# 110 MW Thin-Disk Oscillator

*Semyon Goncharov,* Kilian Fritsch, and Oleg Pronin*

Helmut-Schmidt-Universität / Universität der Bundeswehr Hamburg,
Holstenhofweg 85, D-22043 Hamburg, Germany
E-mail: semyon.goncharov@hsu-hh.de

**Keywords:** ultrafast nonlinear optics; thin-disk oscillator; ytterbium; peak power scaling

**Abstract:** A compact Kerr-lens mode-locked thin-disk oscillator delivering 110 MW output peak power, the highest among all oscillators, is reported. A pulse train with a repetition rate of 14 MHz carries 115 fs long, 14.4 uJ pulses resulting in 202 W of average power. This compact, simple, and stable oscillator is a suitable driver and an important milestone for further high harmonics generation and the development of extreme ultraviolet transportable frequency comb sources.

## 1. Introduction

Recent rapid advances in the development of femtosecond thin-disk oscillators demonstrated a simple mechanism for average and peak power scalability,[1] essentially relying on scaling the mode size inside the Kerr medium (KM) and, thus, opening the path to gigawatt-level peak power directly out of the oscillator. Among the numerous active materials available in the ultrafast field, the Yb:YAG thin-disk is one of the most suitable, in particular for thin-disk operation. As an industrially conventional thin-disk gain medium, it offers 100-200 fs long pulses and relatively high opt.-opt. efficiency, up to 30%, typically resulting in a few hundred watts of output power in the mode-locked regime[1–3] and a few thousand watts in the continuous-wave regime.[4] Particularly, the highest output peak powers demonstrated are 62 MW for a Kerr-lens mode-locked (KLM) oscillator[5] and 66 MW for a SESAM mode-locked oscillator.[6] A KLM oscillator with 102 MW peak power and 50 fs output pulse duration was recently demonstrated.[7] However, the laser operates for only two minutes, what makes it somewhat impractical for applications. Fortunately, these values do not represent a fundamental limitation due to the peak power scaling concept.[1] There is a geometrical procedure behind this concept which implies a simple modification of a cavity resulting in increased mode sizes in the KM without significantly changing any other parameters. This concept provides nearly unlimited scalability of KLM thin-disk oscillators. Moreover, due to its simple design, robustness, and reliability, as well as its ability to utilize the entire gain bandwidth, the KLM Yb:YAG thin-disk oscillator is a favorable laser source for delivering high peak power and femtosecond long pulses with excellent beam quality. Subsequent efficient (up to 95%) nonlinear broadening and pulse compression in multipass cells[8] would result in a 1 GW level laser system delivering sub-15 fs long pulses at a repetition rate of a few megahertz. This oscillator, when CEP stabilized, is an ideal driver to efficiently generate high-order harmonics down to the XUV and deep UV ranges, favoring a coherent optical comb for high precision spectroscopy.[9] There are few spectroscopic applications in the XUV region which are of great interest. For instance, Th-229 is a unique candidate for the development of a new generation of the nuclear clock, which promises to be a few orders of magnitude more precise than the existing atomic-based clock.[10] Even though several groups have measured the isomer transition in Th-229 indirectly,[11] the direct optical measurement remains to be completed. Another appealing application is high precision spectroscopy of singly ionized He atoms to perform tests of QED, for instance, the temporal drift and evolution of fundamental constants.[12]

2. **Experimental results**

In this paper, we report on a compact femtosecond Yb:YAG KLM thin-disk oscillator. This type of oscillator includes an active gain medium, a thin-disk, a set of highly dispersive mirrors (HD), a telescope with the KM in its focus, and a hard aperture (HA). The cavity is usually designed such that the mode size is slightly smaller than the pump spot on the thin-disk. The repetition rate of the cavity can be easily scaled along with the telescope size and varies from a few megahertz[6,13] to a couple of hundred megahertz[14]. Nevertheless, while the upper limit arises from spatial constraints applied to the intra-cavity optics, fundamentally, there is no lower limit dictating the extension of the cavity to at least a few tens of meters, for example, using multipass cells.[6]

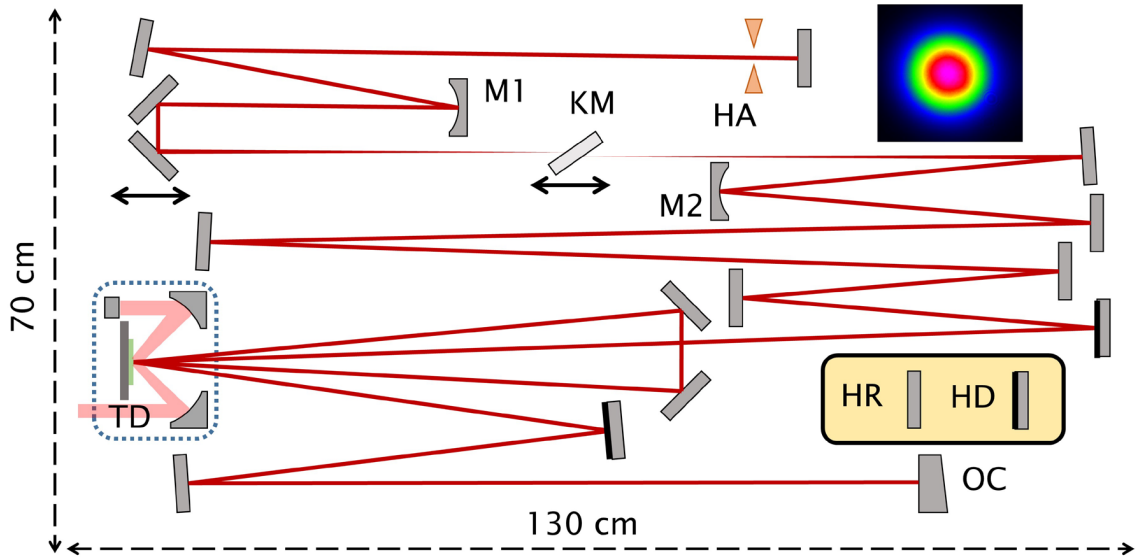

**Figure 1.** Optical layout of the KLM thin-disk oscillator. OC, 19% output coupler; TD, Yb:YAG thin-disk; KM, 5 mm thick crystalline quartz; HA, 5 mm diameter HA; HD, highly dispersive mirror with -2400 fs$^2$ group delay dispersion (GDD); HR, high reflective mirror; M1 and M2, a pair of plano-concave mirrors.

Our thin-disk oscillator design (see **Figure 1**) was similar to a configuration previously reported.[5] A 0.1 mm thick Yb:YAG thin-disk provided by Trumpf served as a gain medium and was pumped by laser diodes at 940 nm. The collimated pump beam was focused onto the thin-disk into a spot diameter of approximately 3.1 mm, achieving a pump intensity of up to 12 kW/cm$^2$ at the full pump power of 900 W and the disk water flow of approximately 5 l/min.

The cavity design was selected in a way such that the average laser mode size across the cavity (defined at the 1/e$^2$ level) was slightly less than the pump spot and maintained a 2.6 mm diameter (see **Figure 2**). In this oscillator configuration, a double-pass geometry was realized (see Figure 1) to increase the overall gain, resulting in a total of 8 passes through the gain medium per round trip. Interestingly, as was demonstrated,[13] a configuration with 16 beam passes through the thin-disk per round trip (or 4 reflections on the thin-disk) is also possible in KLM operation and results in an even higher gain and, thus, a higher output coupling rate of up to 50%.

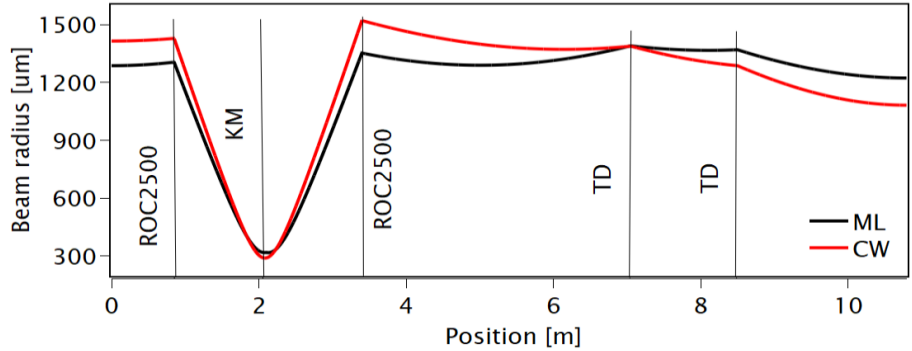

**Figure 2.** Simulated caustics of the cavity in continuous wavelength (CW) and mode-locked (ML) regimes.

Moreover, a cavity design with 52 passes through the active medium allowed an OC rate of 78% in a SESAM mode-locked oscillator.[15] However, when future applications in XUV frequency comb generation are considered, a repetition rate over 10 MHz is highly desirable.

The 5 mm thick Kerr plate was placed in the center of the 2 m long telescopic system consisting of a pair of plano-concave mirrors (see M1 and M2 in Figure 2). Based on the numerical simulation of the beam caustics inside the cavity, the laser mode had a 600 um diameter in the focus of the telescope (see Figure 1). The oscillator ran with both soft and HA mode-locking. The soft aperture was placed in the thin-disk gain medium. The HA was placed close to the end mirror to enhance the overall self-amplitude modulation (SAM) effect and make the mode-locking start-up more reproducible. The HA represented a few-millimeter thick water-cooled copper plate with a 5 mm diameter hole. The Kerr plate placed under Brewster's angle along with the HA and soft aperture in the disk formed a mode-locker that induced sufficient self-phase modulation (SPM) and SAM for the stable pulsed operation.

The formation of a stable soliton pulse in the cavity is possible when the frequency chirp due to SPM is compensated by anomalous GDD. Using this simplified model, a solitary solution can be derived from the nonlinear Schrödinger equation.[16] This solution can exist only for the specified values of peak power $P_{pk}$ and pulse energy $E_p$:

$$\frac{\gamma P_{pk}\tau^2}{|\beta|} = 1 \tag{1}$$

$$E_p = \frac{2|\beta_2|}{\gamma\tau} \tag{2}$$

where γ is the SPM coefficient, τ is the pulse duration, and β is the intra-cavity GDD. A more rigorous analysis can be performed based on a complex Ginzburg-Landau equation,[17] which includes crucial parameters of real oscillators such as SAM, spectral filtering, gain, and loss. A complex interplay between those parameters defines the performance of the KLM regime. In this work, the intra-cavity dispersion management was governed by a pair of HD mirrors, resulting in a round trip GDD of approximately -10,000 fs². Thus, by operating in the anomalous dispersion regime, the oscillator delivered bandwidth-limited sech² soliton pulses. According to Equation 2, the value of the GDD was adjusted to achieve the minimum possible pulse duration. The maximum peak power was set by the resonator configuration. The separation distance between the focusing mirrors M1 and M2 was increased to reach the far edge of the first stability zone. This is a requirement for KLM. The $TEM_{00}$ mode size of the oscillator operating close to the stability edge was much more sensitive to intensity fluctuations introduced by external perturbations. However, by pushing the translation stage with the output-coupler mirror mounted to it, the oscillator was mode-locked, which allowed it to go closer to the center of the stability zone and run in "a stable cavity configuration." The optical cavity was placed inside a monolithic temperature-stabilized aluminum housing in a clean, dust-free environment to enable reliable and stable long-term operation. Importantly, the housing was partially evacuated to provide a low-pressure ambient air environment. It was previously shown[1] that the peak power in thin-disk oscillators can be scaled up by increasing the mode area inside the KM in the focal area of the telescope. Along with the increased intra-cavity peak power, nonlinear effects in ambient air such as self-focusing and SPM become more prominent. These can prevent further increases in peak power or lead to unstable pulses[18] since nonlinear effects in air are spatially distributed (mostly within the Rayleigh range of the telescope) and start to compete with the induced SPM and self-focusing of the KM. Nonetheless, under slight vacuum conditions or in the presence of buffer gas, the impact of ambient air on the total nonlinearity per round trip becomes negligible,[18,19] enabling further efficient peak power scaling. Therefore, at some point in the scaling procedure, it is necessary to operate the oscillator in a low-pressure environment. In the present experiment, the housing had to be evacuated and the residual air pressure in the range of 150 mbar to 600 mbar showed stable mode-locked operation. It was shown[5] that output peak power strongly depends on air pressure. Due to its non-negligible contribution to the overall dispersion, fine-tuning generally leads to an optimal value corresponding to maximum peak power. In this experiment, a residual pressure of 180 mbar was found to be optimal for providing an output peak power of 110 MW.

To initiate the mode-locked operation, the output coupling mirror was disturbed by a magnetic pusher. A specific balance between the induced SPM, SAM, and GDD resulted in the formation of a stable sech$^2$-shaped pulse train. The 10.7 m cavity length defined the pulse repetition rate of 14 MHz. It was beneficial to start the KLM operation when the KM was placed in the focus of the telescope and then the plate was moved out of focus during the mode-locked operation. During this procedure, the output peak and average powers increased proportionally to the pump power. The oscillator configuration was initially pumped at approximately 500 W when the KM was in focus. While shifting the plate outside the focus, the pump power was gradually increased to 900 W. In other words, mode-locking commenced in one configuration where the system easily started, then the system was gradually moved to the optimal configuration with increased SAM depth. Starting the oscillator in this optimal configuration was either not possible or barely reproducible due to the damage to the KM and optics that occurred during oscillator start-up.

There was a limited range of materials that were preferred for utilization as KM's in the experiments. The material needed to have a reasonable nonlinear refractive index as well as high thermal conductivity and low linear absorption for reliable and reproducible high average power KLM operation. In the present experiment, crystalline quartz was used, which fully met these requirements. The choice of KM thickness was based on the following empirical implications. On the one hand, one way to increase induced SPM is to use a relatively thick KM. For such plates, the process of pulse build-up ran smoothly and reliably. However, the thicker the KM, the lower the mode-locking threshold and, thus, the less peak power the oscillator provided. Once the oscillator was mode-locked, the subsequent procedure of moving the thick KM out of focus and the proportional increase of pump power resulted in the appearance of a second or even a third pulse, but not an increase of peak power. On the other hand, for thin Kerr plates, the mode-locking threshold is higher, allowing for high peak power pulses that preserve the single-pulsed regime. However, the mode-locking process became more unpredictable, less reproducible, and frequently led to the damage of the KM and intra-cavity optics. Based on these considerations and previous experience with these types of oscillators, a 5 mm thickness was determined to be optimal for the current resonator configuration. This value ensured both stable, reliable operation and high intra-cavity peak power of 585 MW. The single-pulsed operation was proven by a 1 m long (corresponding to 3.4 ns delay), home-built autocorrelator combined with a fast photodiode with a 175 ps rise time and 2 GHz bandwidth.

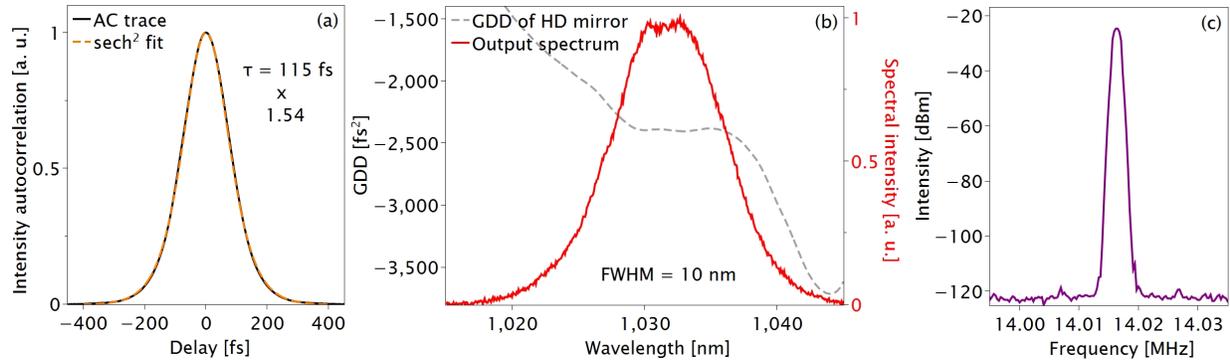

**Figure 3.** (a) Intensity autocorrelation trace of output pulses. (b) Output optical spectrum (red) and GDD curve of HD mirrors (grey). (c) Oscillator repetition rate measured with a radio-frequency spectrum analyzer. The resolution bandwidth is 1 kHz. The peak to background ratio is 95 dB, the sidebands are barely visible.

The 15 ps range was verified by a commercial autocorrelator. The intensity autocorrelation trace, the corresponding optical spectrum, and the RF spectrum of the output pulses are shown in **Figure 3**. The oscillator delivered 115 fs long, 10 nm (FWHM) bandwidth-limited pulses with the time-bandwidth product of approximately 0.32. The output power stability measurement is shown in **Figure 4**. During the first 25 minutes of warming up the oscillator was slowly drifting.

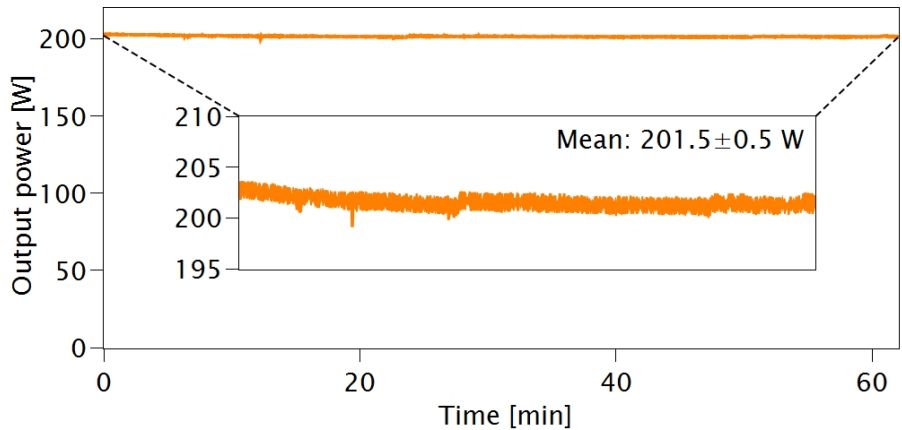

**Figure 4.** Average output power stability measurement. The power fluctuations and drift are attributed to drifts in cooling water temperature.

## 3. Discussion

This record-high output peak power of 110 MW was achieved mainly due to the following reasons. The beam size in the KM was enlarged by extending the telescope to 2.5 m, whereas the previous work used a 2.0 m long telescope. After this procedure, the intra-cavity peak power was scaled from 420 MW to 585 MW, according to the previous experimentally demonstrated geometric peak power scaling concept.[1] This provided additional verification of the applicability of this concept. Moreover, in the current experiment, the pump power density on the thin-disk was increased from 9 kW/cm$^2$ to 12 kW/cm$^2$ by increasing the overall gain, making it possible to operate the oscillator with a higher output coupling rate of 19% compared to 15%[5] without sacrificing a pulse duration. These experimental steps resulted in a 110 MW peak power and 202 W average power directly from the oscillator.

Taking into account the results of this experiment, no fundamental limits were observed for further scaling of KLM thin-disk oscillators towards the gigawatt level of output peak power. A simple geometrical peak power scaling concept should be followed by increasing the area of the KM placed in the focus of the telescope. As was demonstrated in this work, the procedure can be realized by implementing a 4f imaging telescope while at the same time proportionally adjusting the intra-cavity dispersion according to Equations 1 and 2.

Interestingly, the oscillator spectrum was very close to the emission bandwidth of the Yb:YAG gain medium, which is about 9 nm FHWM. In principle, it is possible to overcome this limitation by further increasing the SAM using a distributed KLM technique,[20] or in a way it was recently shown in[7] which already demonstrated 50 fs long output pulses directly from a Yb:YAG thin-disk oscillator.

As previously mentioned, the average pump power density in the oscillator reached 12 kW/cm$^2$, which was close to the critical value when thermal lens distortions in the disk might get pronounced. Those effects typically change the stability of the cavity.[21] Depending on the resonator configuration, they may even favor the mode-locked operation, for instance, by pushing the cavity towards the stability edge due to the induced thermal lens. Thereby, one of the ways to avoid such unpredictable scenarios is to maintain a pump power density for the thin-disk by enlarging the pump spot, which is also a key to further average power scaling. Further scaling can also be accomplished by increasing the number of passes through the disk. Both approaches were previously demonstrated.[5,13,22]

Furthermore, because of the fundamental soliton mode-locking regime and its low noise characteristics,[23,24] CEP stabilization of this type of oscillator no longer presents a great challenge. Moreover, the stabilization performance can be further improved if technical noises such as flicker noise can be overcome, which may be drastically suppressed by minimizing the repetition rate and maximizing the intra-cavity peak power.[25] This is precisely the case with the oscillator in this experiment; thus, this oscillator is expected to be CEP stabilizable with excellent short and long-term performance.

To ensure a good conversion efficiency of $10^{-7}$–$10^{-6}$ in the upcoming experiments on direct deep UV and XUV generation in a gas jet, a high peak power laser favoring ultrashort pulses of a few tens of femtoseconds would be required.[26,27] Both requirements can be fulfilled in a single step with nonlinear broadening and pulse compression in a multipass cell.[28,29] Depending on the configuration, for a cell filled with nonlinear gas, a compression factor of no more than 10,[30,31] and possibly even more than 10,[32] can be achieved, preserving a relatively clean shape of the pulse with over 80% power in the main peak.[29,31] Moreover, due to their simplicity, robustness, and high transmission (up to 95 %), these multipass cells can be stacked consecutively to yield an overall broadening and compression factor greater than 20 with approximately 1 GW, sub-10 fs pulses.

4. **Conclusion**

We have demonstrated a compact table-top Yb:YAG femtosecond thin-disk oscillator delivering 110 MW output peak power. The spectral FWHM of the generated output pulses was close to the emission bandwidth of the Yb:YAG gain medium and resulted in 115 fs long soliton pulses containing 14.4 uJ of energy directly from the oscillator without any external amplification. Additionally, we have experimentally shown that the concept of geometric power scaling [1] can be extended to even higher output peak powers. A subsequent spectral broadening and pulse compression of oscillator pulses down to sub-10 fs in gas-filled multipass cells is feasible with the upcoming experiments, resulting in a gigawatt-level amplification-free laser system. As shown previously, CEP stabilization of this type of oscillator with an intra-cavity acousto-optic modulator is possible with a performance at levels as low as 90 mrad of integrated phase noise.[24] Even better values are expected for the current system. Therefore, this compact and simple oscillator is a promising driver for XUV frequency combs and consequent high precision XUV spectroscopy experiments.

**Acknowledgments**

Considering the difficulty in setting up the new professorship in combination with typical university bureaucratic procedures, we would like to sincerely acknowledge a few facilitators: A. Borchers, D. Kiesewetter, and A. Puckhaber.